# System Level Analysis and Management of Orbital Debris Using Empirical Dynamic Modeling

Asaad S. Abdul-Hamid,[*] and Hao Chen[†]
*Stevens Institute of Technology, Hoboken, New Jersey, 07030, United States*

**Orbital debris is a pressing problem which presents a danger to global space operations and a barrier to continued development of the space economy and space infrastructure. As research continues regarding orbital debris, there is a need for tools to understand the system-level implications of orbital debris solutions. This research considers the orbital debris problem as a dynamic process. Based on dynamic system theories, time-series variables of the numbers of orbital debris, orbital objects, and object launches should be causally linked, which means they share a common system attractor manifold. We propose a data-driven method based on complexity science to reconstruct a shadow attractor of the dynamic system using limited observable variables. The reconstructed shadow attractor helps us to understand the fundamental system dynamics for orbital debris and enables us to simulate the future of the orbital debris system based on changes to policy. These findings represent a significant advancement in our ability to understand high level impacts of space system policy with limited data available.**

## Nomenclature

$\rho$ = correlation or prediction skill

$\theta$ = S-map Localization parameter

$E$ = Embedding Dimension

$\tau$ = Time delay

$d$ = Manhattan distance

This paper is a substantially revised version of the paper AIAA 2024-2053, presented at the 2024 AIAA SciTech Forum, Orlando, FL, January 8-12, 2024.
[*] PhD Candidate, Department of Systems Engineering, AIAA student member.
[†] Assistant Professor, Department of Systems Engineering, hao.chen@stevens.edu, AIAA senior member. (Corresponding Author)

# I. Introduction

Earth's orbit has become significantly populated by defunct satellites, old vehicle stages, and fragments from explosions or collisions that happened in space. These all contribute to space junk, otherwise known as orbital debris. As of 2022, there are over 15,000 orbital debris objects represented in NASA's LEGEND orbital debris model for Lower Earth Orbit (LEO) [1], as seen in Fig. 1. This debris presents a real and present danger to current and future operations, as a collision between active satellites and the orbital debris could be catastrophic to the operation of the satellite. The fragments from such a collision would also increase the total number of orbital debris [2]. Lethal non-trackables are orbital debris objects that are too small to be reliably tracked but still have enough kinetic energy to cause significant damage to any operational spacecraft or satellite that they come into contact with [3]. This is a problem that will only grow with time. As many major players have emerged from the private sector, commercial space activity has stimulated a huge boom in space operations, especially pertaining to the deployment and use of satellites in LEO [4-5]. In order for successful operations in space to continue, orbital debris must be mitigated.

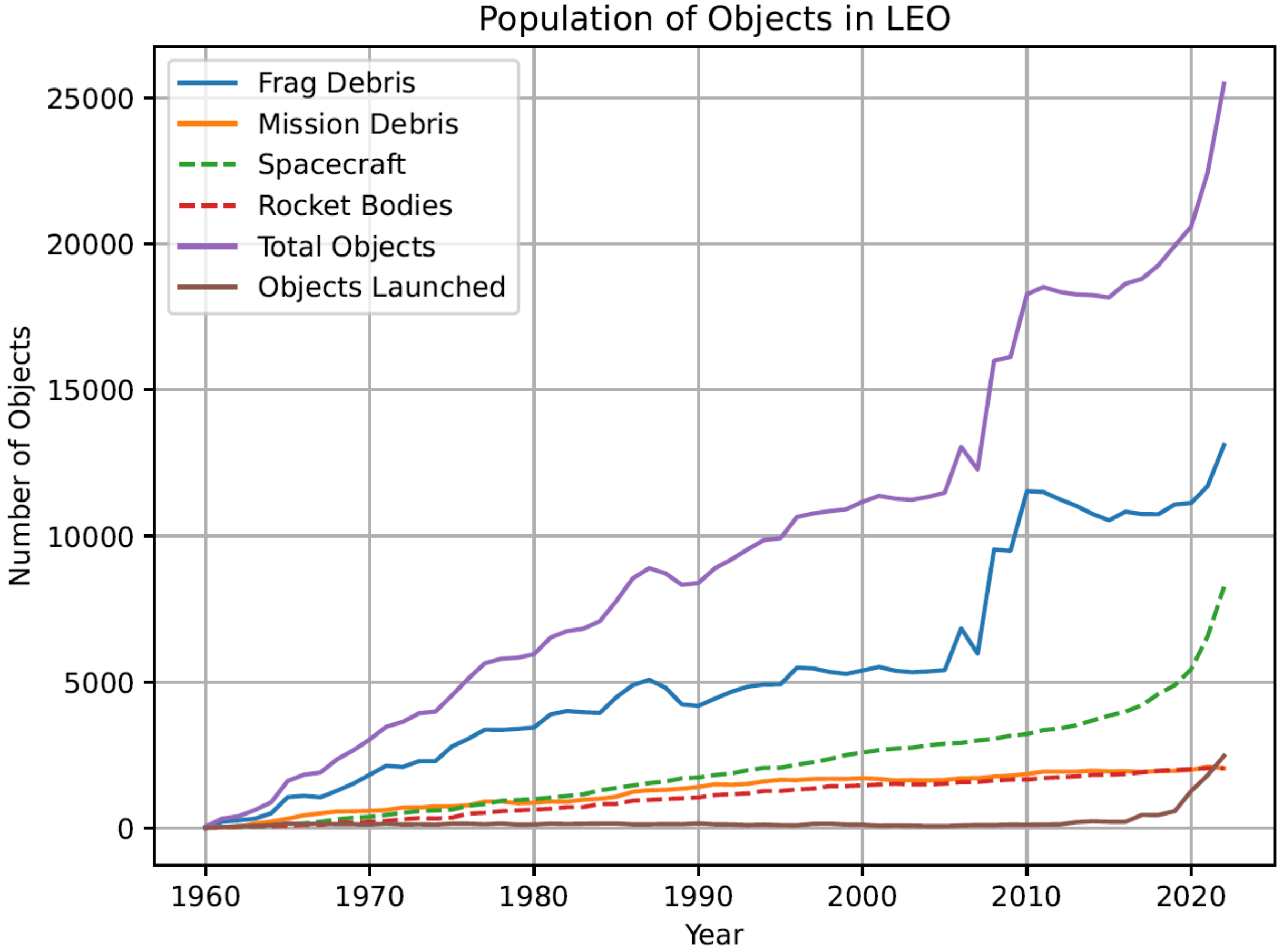

**Fig. 1 Population of Objects in LEO from 1960 to 2022 (data extracted from Ref. [1]).**

The literature regarding the orbital debris problem can be broken down into three major areas: policy, technical, and economic. Policy discussion surrounding orbital debris is based on the many different national and international guidelines regarding orbital debris. As the discussion about the imminent danger of orbital debris has increased over the years, national and multinational entities have realized that leaving a satellite in orbit for multiple decades is not sustainable for maintaining an operable orbital environment [6]. The development of a comprehensive global international policy for mitigating space debris is a topic of great importance in this area. Previously released guidelines have been loosely complied with, and critically, there is no system of liability which establishes who and to what degree is responsible for cleaning up orbital debris [7]. This is a natural result of the fact that space does not belong to any particular nation or even group of nations. It is a common domain to be experienced by anyone who can access it [8].

The technical aspect of the literature is comprised of characterizing and tracking the debris, accurately modeling the debris environment, and various methods to perform active debris remediation, sometimes called active debris removal [9]. Modeling of the orbital debris environment for both current understanding and forecasting purposes has been a longstanding effort by public space entities. For instance, NASA maintains their Orbital Debris Engineering Model (ORDEM) and the LEGEND model for characterization, tracking, and forecasting of the debris environment [10]. On the matter of active debris remediation, the literature proposes a number of approaches for removing the debris that currently exists in Earth's orbit. This includes tugging debris into lower orbits for reentry into Earth's atmosphere, using lasers, ion beams, or small satellites to nudge debris to avoid collisions, and the use of orbital sweepers to intercept small pieces of debris [11].

The economic portion of the literature largely deals with the cost efficiency of active debris remediation missions. A major point of interest in this field is the optimal number or mass of targets to be addressed in each mission. In 2014, Braun et al. devised a cost estimation for the removal of multiple debris objects depending on their priority level [12]. More recently, in 2023, Colvin et al. from NASA released a cost-benefit analysis which detailed the price and impact of removing either the 50 highest priority targets in LEO or removing 100,000 pieces of small debris [11]. These cost models are a significant effort toward comprehensively estimating remediation costs, but since an active debris remediation mission has never been launched, these estimates are based on assumptions that may be disproved in practice. These cost estimation models and many economic factors of active debris remediation are likely to see a major update after 2026, when the company ClearSpace plans to launch the first mission to remediate a piece of debris [13]. This is a mission sponsored by the European Space Agency, and will be a major milestone in space operations and the mitigation of orbital debris.

A more recent aspect of the discussion about space debris is a turn towards systems thinking to characterize the current situation and develop solutions. Systems thinking is a relatively new discipline, which does not just focus on the particular mechanics of a singular system like the orbital debris environment or the plethora of satellites in Earth's orbit. Instead, systems thinking views individual components as part of a larger interconnected network, examining the way the components interact with each other, and optimizing their performance as part of the bigger picture. Verma et al [14] applied this method to the problem of orbital debris, and in this framing of the problem, have been able to identify policy weak points, and suggest changes to address them.

This paper seeks to add to the analysis of space debris through systems thinking by considering the problem as a dynamic process to analyze the interconnected systems that contribute to the problem of orbital debris. We will create a data-driven modeling method that recovers high-level system information without requiring prior understanding of the system's mechanics. This modeling method will be able to predict the future growth of the orbital debris population and simulate how policy changes might affect that growth trajectory. Central to this capability is Takens' Embedding Theorem [15], which shows that by looking at only one system variable, one can recover information about all the other variables in the system. It has been used for a number of applications, but has been notably discussed within the ecological community [16-18]. Ecological systems are often measured via a large breadth of data over a relatively short period of time, and the systems themselves are complex, nonlinear, and dynamic. The empirical dynamic modeling (EDM) method developed based on the Takens' Embedding Theorem is likely a natural fit for these systems then, as it is capable of recovering essential system dynamics without needing an extended amount of time series data. In fact, the greater the breadth of the data, meaning the more system variables provided for a particular portion of time, the better EDM is at reconstructing the system mechanics. The orbital debris problem can similarly benefit from Takens' Embedding Theorem. When examining annual debris activity over the entire history of human space activity, one arrives at just over 60 data points. Additionally, the history of non-cooperating and cooperating missions in space makes the system complex, as there are entities with their own interests operating within the system. These factors make EDM an attractive option for analyzing an orbital debris system which is complex and dynamic. The use of dynamic modeling for the orbital debris problem is novel, and will provide new insight and further potential in understanding orbital debris and other space systems. The insights from this analysis will expand the space community's capability of attaining a holistic and robust understanding of the contributing factors to orbital debris, in addition to being able to quantitatively predict how multiple space systems will evolve as orbital debris is addressed.

The remainder of this paper is organized as follows. Section II first introduces the dynamic modeling for the orbital debris problem and proposes the system state reconstruction method based on Takens' Embedding Theorem. In Section III, we exhibit the forecasting ability and conduct simulations to give a quantitative basis for how new policy might mitigate the orbital debris population. Finally, Sec. IV concludes the paper and discusses future works.

## II. Methodology

We consider orbital debris as a dynamic system. Suppose the $\boldsymbol{N}$-dimension system states $\boldsymbol{Y}(\boldsymbol{t}) \in \mathbb{R}^{\boldsymbol{N}}$ evolve following a differential function $\boldsymbol{\Psi}: \mathbb{R}^{\boldsymbol{N}} \rightarrow \mathbb{R}^{\boldsymbol{N}}$, written as

$$\dot{\boldsymbol{Y}} = \Psi(\boldsymbol{Y}) \tag{1}$$

For the complex system, only some of the system attributes are measurable, directly or indirectly [19-21]. We define all measurable system state variables as $\boldsymbol{X}(\boldsymbol{t}) \in \mathbb{R}^{M}$, where M is typically much smaller than the original system state dimension $\boldsymbol{N}$ [22]. Using traditional parametric equation-based methods, we can create the measurement function $\boldsymbol{X}(\boldsymbol{t}) = \Phi(\boldsymbol{Y}(t))$ to connect measures with original system states, where $\Phi: \mathbb{R}^{\boldsymbol{N}} \rightarrow \mathbb{R}^{M}$, such as through surrogate modeling (e.g, polynomials, splines, and kriging) [23-28] or physical-based modeling [29-31]. Solving these problems using conventional parametric approaches brings two main challenges, as shown in Fig. 2. First, we need to obtain accurate measurement functions, $\Phi$, that pose dual risks of model misspecification [32] and variable misidentification [33, 34]. Even if we know the correct form of measurement functions, fitting the parametric models can also be problematic due to limited data and system nonlinearity. A complex parametric model is often accompanied by strong assumptions and intensive data requirements [35]. The second challenge is the potential nonequilibrium dynamics in system state evolvement, such as the satellite collision as shown in Fig. 1. More importantly, the influence can be bidirectional and state-dependent which cannot be effectively captured by equilibrium-based models through hypothesized rules. In response to these challenges, we propose a data-driven framework with minimal system behavior assumptions for debris system state identification and robust forecasting.

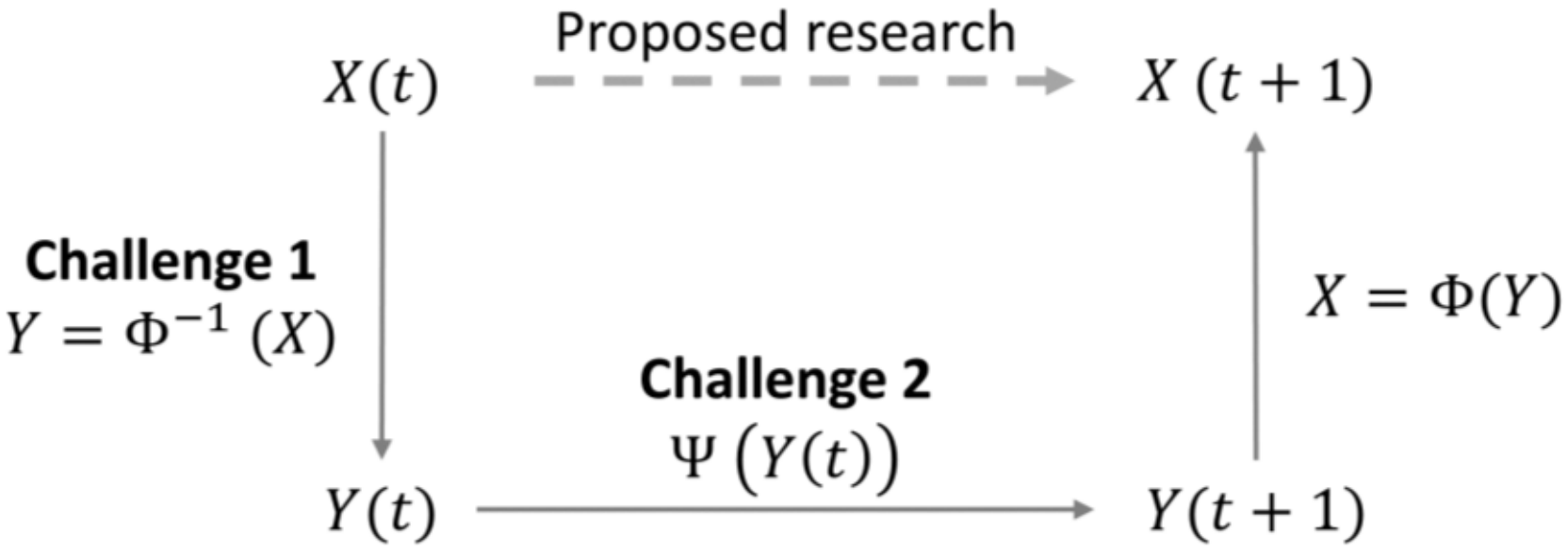

**Fig. 2 Challenges in orbital debris forecasting.**

### A. Attractor Reconstruction

Instead of attempting to achieve a comprehensive understanding of the system behavior using numerous assumptions to approximate the measurement function Φ and dynamic evolvement function Ψ based on hypothesized rules, the EDM method relies on time series monitoring data to reveal the system dynamics between variables. Collectively, system states and their trajectories form an attractor manifold $\mathcal{M}$. They form the original attractor state space, as shown at the top of Fig. 3.

The debris monitoring data represents an observation of the original dynamic system attractor. Mathematical results prove that generically, information about the entire system attractor is contained in any one of the observation variables [15, 36]. Therefore, we can leverage proxy variables (i.e., $X_1(t)$) and delay coordinate mapping to reconstruct a shadow manifold, as shown at the bottom of Fig. 3. With this shadow manifold, we can determine the relative contributions of contextual factors to orbital debris state evolvement. When parametric models are required for orbital debris modeling, this shadow manifold can also play a complementary role in causal analysis, variable identification, and the construction of reliable equation-based models [37].

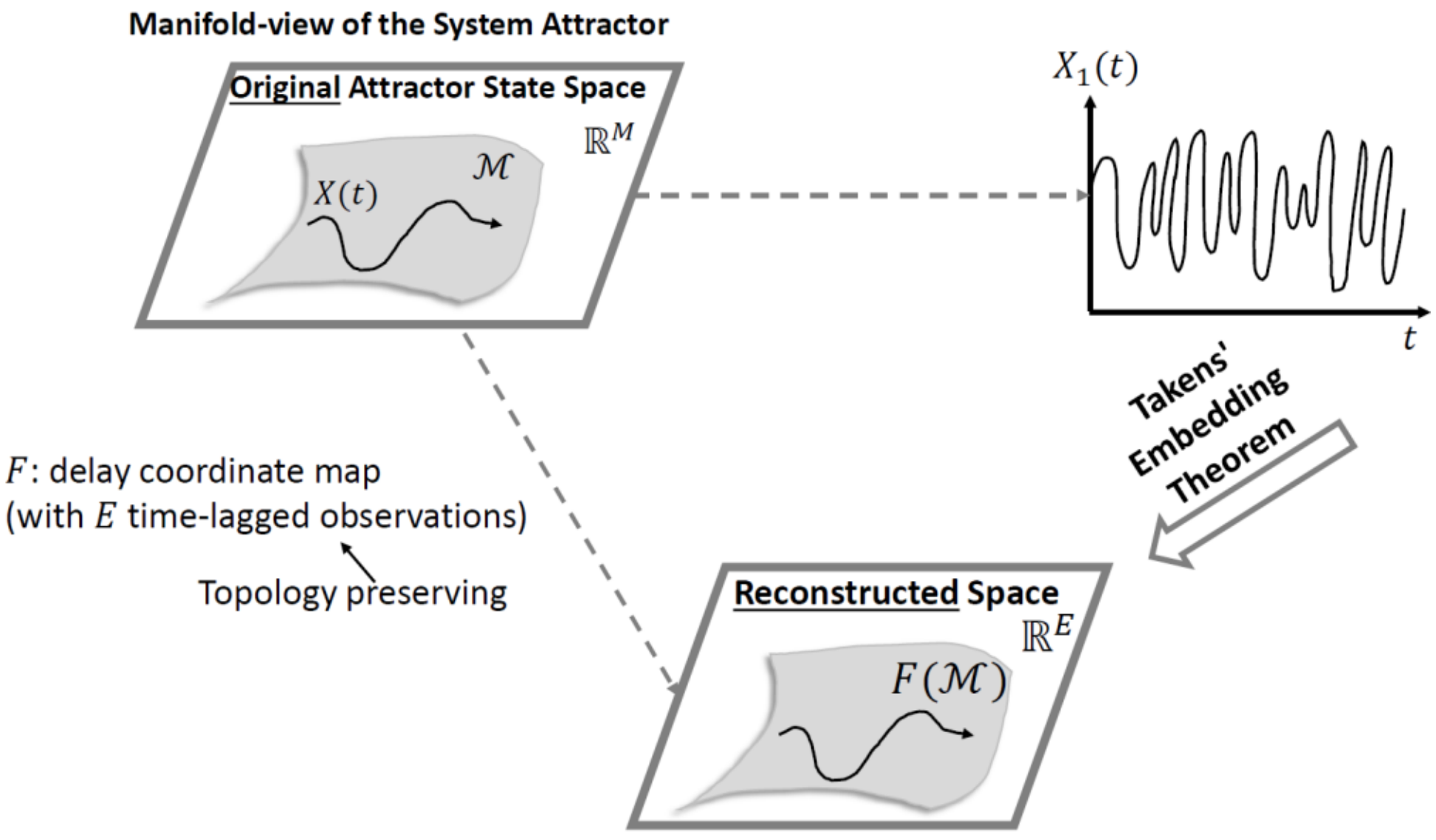

**Fig. 3 EDM Concept.**

The reconstruction of system dynamics from time series monitoring data is critical to our analysis. With $E$ time-lagged observations and time step $\tau$ , we have system monitoring data $\{X(t), \ldots, X(t-(E-1)\tau)\}$. Based on Takens' embedding theorem, if $E > 2M$, where $M$ is the dimension of the attractor, we can reconstruct a shadow manifold that preserves the topology of $\mathcal{M}$ through a delay coordinate map $F_{(\Phi,\tau,E)} : \mathcal{M} \rightarrow \mathbb{R}^{\boldsymbol{E}}$ [15,19]. The unknown or unobserved variables are substituted by the lags of monitoring time series data. Generalizations of Takens' theorem allow us to reconstruct the shadow attractor using multiple time series [38]. For example, we can reconstruct the state space based on the monitoring data of $\{X_1(t), X_2(t), X_3(t)\}$. One possible reconstruction vector is $\langle X_1(t), X_1(t-\tau), X_2(t), X_2(t-\tau), X_3(t), X_3(t-\tau)\rangle$.

**B. Simplex Projection for Forecasting**

In this section, we will discuss prediction via Simplex Projection. Simplex projection has been utilized in the ecological community to forecast system states by using only chaotic time series information [16,17]. With the information in the reconstructed shadow manifold, we can forecast the future state of every target point on the manifold. However, the delay coordinate map $F(\mathcal{M})$ can be considered as an embedding of the original attractor manifold $\mathcal{M}$ onto the reconstructed space for almost every choice of the measurement function, $X(t) = \Phi(Y(t))$ [15,19]. For an interested debris system measure, $X_1(t)$, we identify $b$ closest neighbor points on the shadow manifold $F(\mathcal{M})$. These $b$ neighbors form a bounded simplex that surrounds $F_{(\Phi,\tau,E)}(Y(t))$ in the $E$-dimension space. Then, we can project this simplex to the future, and the forecasted value of $X_1(t + 1)$ is the weighted average of projected $b$ neighbors. The weights are determined by the Euclidean distance between $X_1(t)$ and the neighbor, indexed by $n(j)$ for the $j$-th closest neighbor. Then, the weight for neighbor $j$ can be written as:

$$w_j = exp\left(-\frac{\|X_1(t) - X_{1,n(j)}(t)\|}{\|X_1(t) - X_{1,n(1)}(t)\|}\right) \quad (2)$$

Simplex Projection is also critical for understanding the dimensionality of the attractor. Dimensionality is represented by *E*, the embedding dimension. This is the minimum number of dimensions necessary for representing the attractor and enabling accurate predictions. For instance, an *E* value of 2 would mean $X_1(t)$ and $X_1(t-1)$ are sufficient time series for reconstructing the state space, assuming a $\tau$ value of 1. Predictions from Simplex Projection are verified for accuracy using the Pearson correlation coefficient between observations and predictions ρ, as well as Root Mean Squared Error (RMSE). A ρ of 1 would indicate 100% perfect accuracy, and a low RMSE relative to the scale of data being analyzed is preferred. $E$ and $\tau$ are hyperparameters which must be optimized to maximize the prediction correlation $\rho$ and minimize the RMSE.

**C. Convergent Cross Mapping for Causality Detection**

Convergent Cross Mapping (CCM) enables verification of causality between system variables. Consider that the previous example system represented by attractor manifold $\mathcal{M}$ is comprised of time series $X_1(t)$, $X_2(t)$, and $X_3(t)$. How can we be sure that these collections of time series belong to the same system? In 2012, Sugihara et al [39] proposed a method to confirm if two variables are causally related using Takens' Embedding Theorem and state space reconstruction. The CCM test for proving that $X_1$ causes $X_2$ would be as follows: Using time delays of system function $X_2(t)$ builds a shadow manifold $F_{X_2}(\mathcal{M})$ which corresponds directly with the original attractor manifold $\mathcal{M}$. The information being reflected in the shadow manifold $F_{X_2}(\mathcal{M})$ would also contain information about the original system variables, such as an analog for the original $X_1(t)$. By using Simplex Projection, that reconstructed state space can be used to forecast the evolvement of the system. This is referred to as cross-mapping, using the state space reconstruction of one system variable to predict another system variable. If the correlation ρ between the prediction of $X_1(t)$ and the original system variable $X_1(t)$ converges at a positive value over time, then it is said that $X_1$ causes $X_2$ . This is because the CCM test verified that information about $X_1$ could be recovered from the state space reconstruction which only used the $X_2$ variable, and that amount of information increased over time. Specifically, CCM looks for the signature of $X_1$ in $X_2$'s time series data. The strength of that causative effect can also be measured using this method. Testing the reverse causality would require this same process, but with the roles of $X_1$ and $X_2$ reversed. If the neighbor points are indexed by $n$, we can write the monitoring data for $X_{2,n}(t)$ as

$$X_2(t_n) = ( X_2(t_n), \dots , X_2(t_n - (E-1)\tau) = \left(X_{2,n}{}^{(1)}, \dots , X_{2,n}{}^{(E)}\right) \tag{3}$$

Then, the cross-mapping equation for detecting the signature of $X_1(t)$ in the state space of $X_2(t)$ is as follows:

$$signature_{X_1}\left(X_2(t)\right) = \sum_{n=1}^{E+1} \frac{e^{-d\left(X_2(t), X_2(t_n)\right)}}{\sum_n e^{-d\left(X_2(t), X_2(t_n)\right)}} \left( \frac{X_{2,n}{}^{(2)} - X_{2,n}{}^{(1)}}{X_{2,n}{}^{(1)}}, \dots, \frac{X_{2,n}{}^{(E)} - X_{2,n}{}^{(E-1)}}{X_{2,n}{}^{(E-1)}} \right) \tag{4}$$

where $E$ is the total number of data points, which is also the dimension of reconstructed state space; $d$ is the Manhattan distance. This signature of $X_1(t)$ from the reconstruction of $X_2(t)$ is then compared with the real values of $X_1(t)$ to test if the Pearson correlation ρ converges at a positive value.

**D. S-Mapping for Nonlinearity Analysis**

Where Simplex Projection looks at only the nearest neighbors, S-Mapping (Sequential locally weighted global linear mapping) looks at the entire data set, and weighs points in the data set differently. We use S-Mapping to analyze the nonlinearity of multiple interested debris system measures $X(t)$ [32, 40]. Whereas Simplex Projection relies on the weighted average values of the neighbor points to perform forecasting, S-Mapping is a locally weighted linear regression. In S-Mapping, each point in the dataset, or library, is assigned a weight. For a debris system measure $X_1(t)$, the weight of its $j$-th closest neighbor can be written as,

$$w_j = ex\,p\left( -\theta \frac{\left\|X_1(t) - X_{1,n(j)}(t)\right\|}{\frac{1}{Q}\sum_{q=1}^{Q}\left\|X_1(t) - X_{1,n(q)}(t)\right\|} \right) \tag{5}$$

where $\theta \geq 0$ is the nonlinear parameter and $Q$ is the size of the data set. $\theta$ shows the importance of local system states in system dynamics forecasting. When $\theta > 0$, the S-Map more strongly considers the state of the point in question, as opposed to weighing all points equally if $\theta = 0$. S-Mapping is similarly useful for analyzing the nonlinearity of a system. When the localization parameter $\theta = 0$, the model is linear autoregressive, as opposed to being nonlinear when $\theta > 0$. Thus, a system's linearity can be confirmed by comparing the prediction skill ρ when $\theta = 0$ and when $\theta > 0$. If the highest prediction skill ρ occurs when $\theta > 0$, a system is nonlinear and state-dependent.

While S-Mapping is a nonparametric form of analysis, meaning that it makes minimal assumptions regarding data distribution, it can be used to gain quantitative information about how system variables interact. S-Mapping shows the partial derivative of each system variable with respect to the variable in question, where the partial derivative estimates are the local regression coefficients. For instance, we could use S-Mapping to predict the $X_1(t)$ function in our $\mathcal{M}$ attractor manifold. In addition to predictions of $X_1(t)$'s time series data, it would produce the partial derivatives of $X_1$ with respect to the other system variables, $\frac{\partial X_1(t)}{\partial X_2(t-\tau)}$ and $\frac{\partial X_1(t)}{\partial X_3(t-\tau)}$, for each observed point and each predicted data point. This can be used to track the strength and sign of interactions between variables.

## III. Results

### A. Data Source

We will analyze a system comprising the annual data for debris objects in LEO, number of objects launched into space, and total objects in LEO. These variables will be represented by X, Y, and Z, respectively. The time series data on debris objects and total objects were aggregate population data from NASA's LEGEND model [1], and the objects launched were publicly available via ourworldindata.org, which lists the original source as being the United Nations Office for Outer Space Affairs [41]. The scope of this paper will pertain only to LEO, including debris objects with an altitude of 2,000 km or less. Only objects with characteristic length greater than 10 cm are considered.

EDM calculations will be done using the pyEDM package in Python [42]. For the time lag parameter $\tau$ we will utilize the pyEDM default value of 1, which has a unit of years since our data is yearly. The software package also allows selection of the exclusion radius parameter. This parameter ensures that time indices near a prediction point are not considered during forecasting, aiding the analysis of true system dynamics and not simple autocorrelation [43]. For our EDM analysis, we designate the exclusion radius to be the product of the embedding dimension and the time delay, $E \cdot \tau$. While EDM can be used to handle unevenly spaced sets of empirical data [44], pyEDM requires that data be of equal intervals, so the range of all data in our analysis is 1960 to 2022.

### B. Simplex Projection

To begin our analysis, we will use univariate Simplex Projection to predict the future of debris objects in LEO. This prediction will only be done using univariate Simplex Projection, as the results of the optimal embedding dimension will be necessary for the CCM analysis. After running multiple trial simulations to find the optimal embedding dimension E, we can use Simplex Projection to predict the debris population up to 2050, as seen in Fig. 4.

An embedding dimension of E = 4 is the most optimal as it provides the highest prediction skill ρ. The model is successively fed each year of historical data from 1960 to 2022. From 1991 to 2022, the model predicts alongside the historical data, interpolating from the known data set and measuring its ability to match the true historical data as the prediction skill ρ. After 2022, the model is purely predictive, extrapolating for predictions outside of the observed data. The historical data serves to provide information about system dynamics which can be recovered via EDM. These dynamics implicitly include factors contributing to debris propagation, such as collisions with debris or active spacecraft in LEO, and explosions.

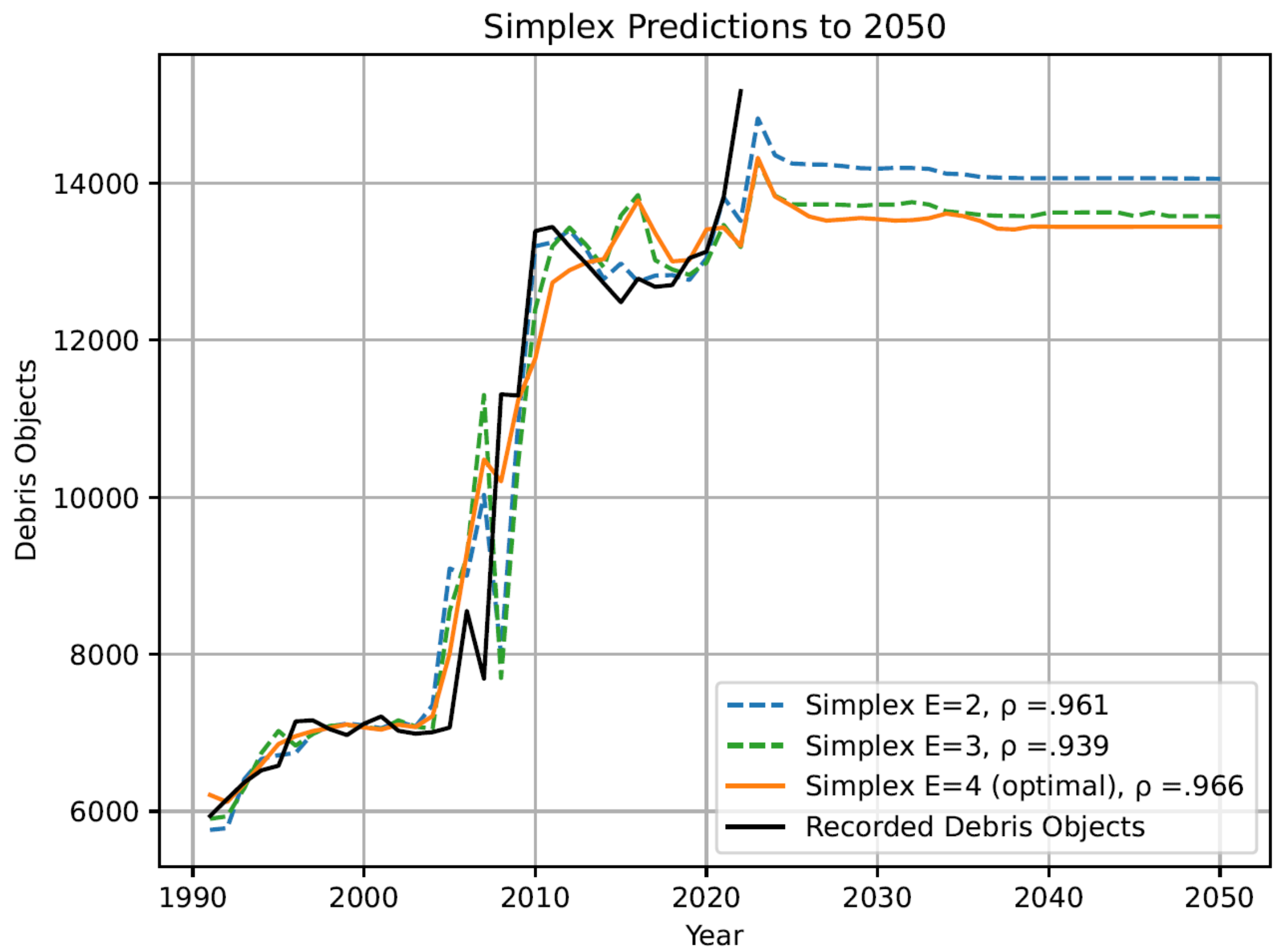

**Fig. 4 Simplex Projection forecast out to 2050, with multiple embedding dimensions E.**

While the optimal Simplex Projection at E=4 has a prediction skill of 96.6%, and follows relatively closely to the original recorded data, the flattening out of the debris population in forecasted data is unrealistic. The debris population is capable of self-propagation due to catastrophic collisions, and thus is only likely to increase into the future [2]. The constant line exhibited by all forecasts indicates that Simplex Projection, where only nearest neighbors are considered, is a poor fit for forecasting the orbital debris environment.

### C. Convergent Cross Mapping

Using the optimal embedding dimension of E=4 from the univariate Simplex Projection, the next analysis is CCM. We will test if our assumed system variables: the number of debris in LEO, number of objects launched into space, and total objects in LEO, are causally related. Results can be seen in Fig. 5. (A) to (C) are CCM of debris vs. objects launched (X:Y), debris vs. total bodies (X:Z), and objects launched vs. total bodies (Y:Z).

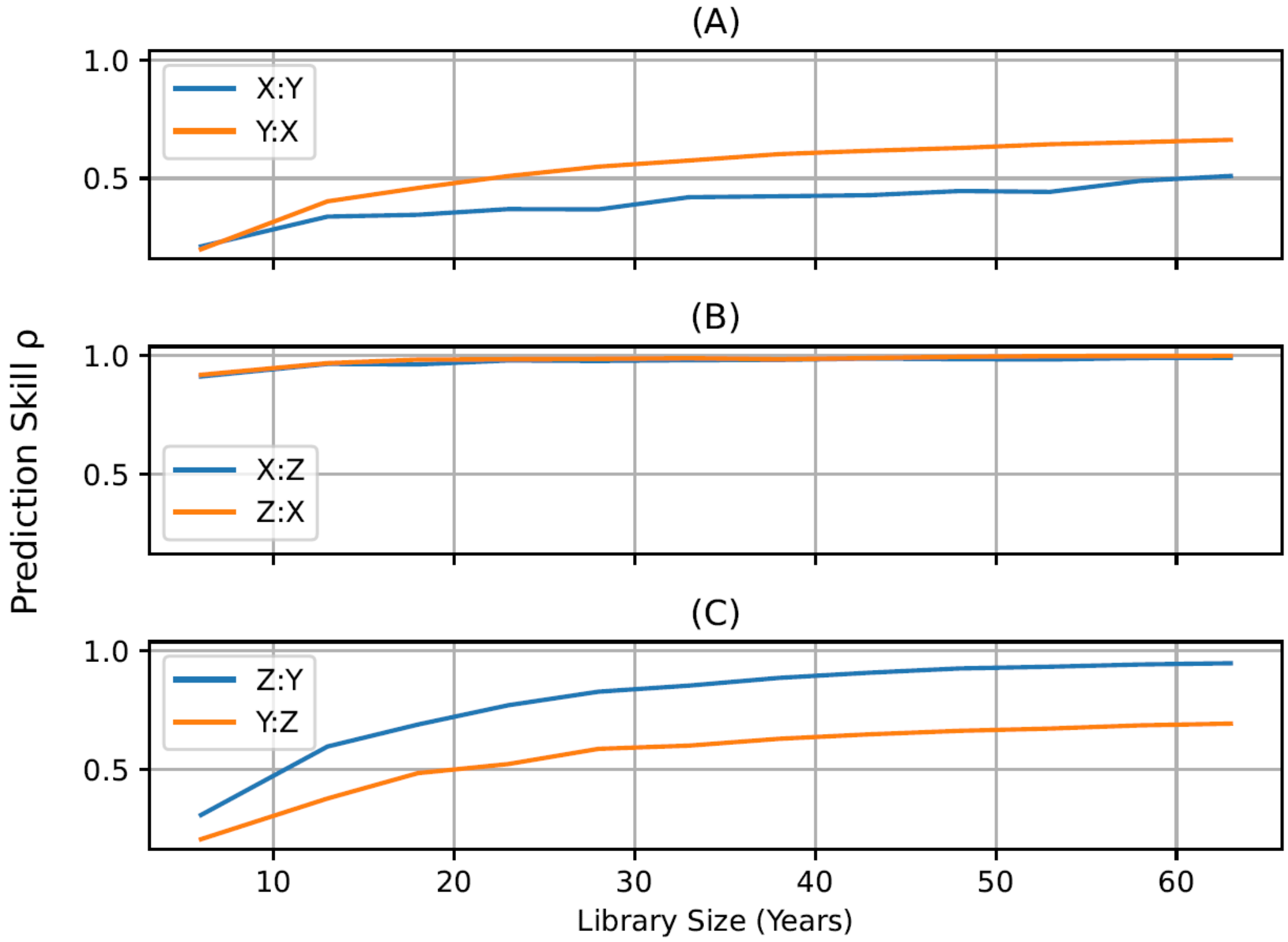

**Fig. 5 Convergent Cross Mapping to verify causality between variables.**

In all cases, the correlation ρ increases as the library size increases, but only graph B exhibits convergence toward a common value for both variable pairings. Thus, the CCM analysis supports that the debris objects in LEO and the total bodies in LEO are causally related, exhibiting bidirectional causality. The results from other variable pairings also converge to a positive value, but with a lower prediction skill than in graph B. This indicates that causality is strongest between X and Z variables, so predictions using those variables will be more precise.

### D. S-Mapping

Since Simplex Projection is an ill fit for our data, we will move to the more advanced forecasting method, S-Mapping. We continue to use the optimal embedding dimension E=4 as a parameter for S-mapping. Similar to Simplex, we need to find the optimal parameter θ, which is the localization parameter. $\theta = 7$ is the most optimal for S-Map using X and Z variables, as it provides the highest prediction skill, with a ρ value of 97.1%. The weighted consideration of all points in the library with S-Mapping, as opposed to just the nearest neighbors with Simplex Projection, should result in a more accurate prediction of future behaviors. Figure 6 shows that S-Mapping predicts an increase in the debris environment, with over 30,000 debris objects > 10 cm being predicted by 2050. Each prediction features a 95% confidence band based on the variance at each prediction point. There is an increase in uncertainty in prediction between 2003 and 2013, which can likely be attributed to new dynamics being added into the system in this time period. Such dynamics would include the 2007 ASAT test and the 2009 satellite collision. Otherwise, the uncertainty increases as the prediction horizon increases. The prediction using the most optimal θ parameter also experiences the least uncertainty.

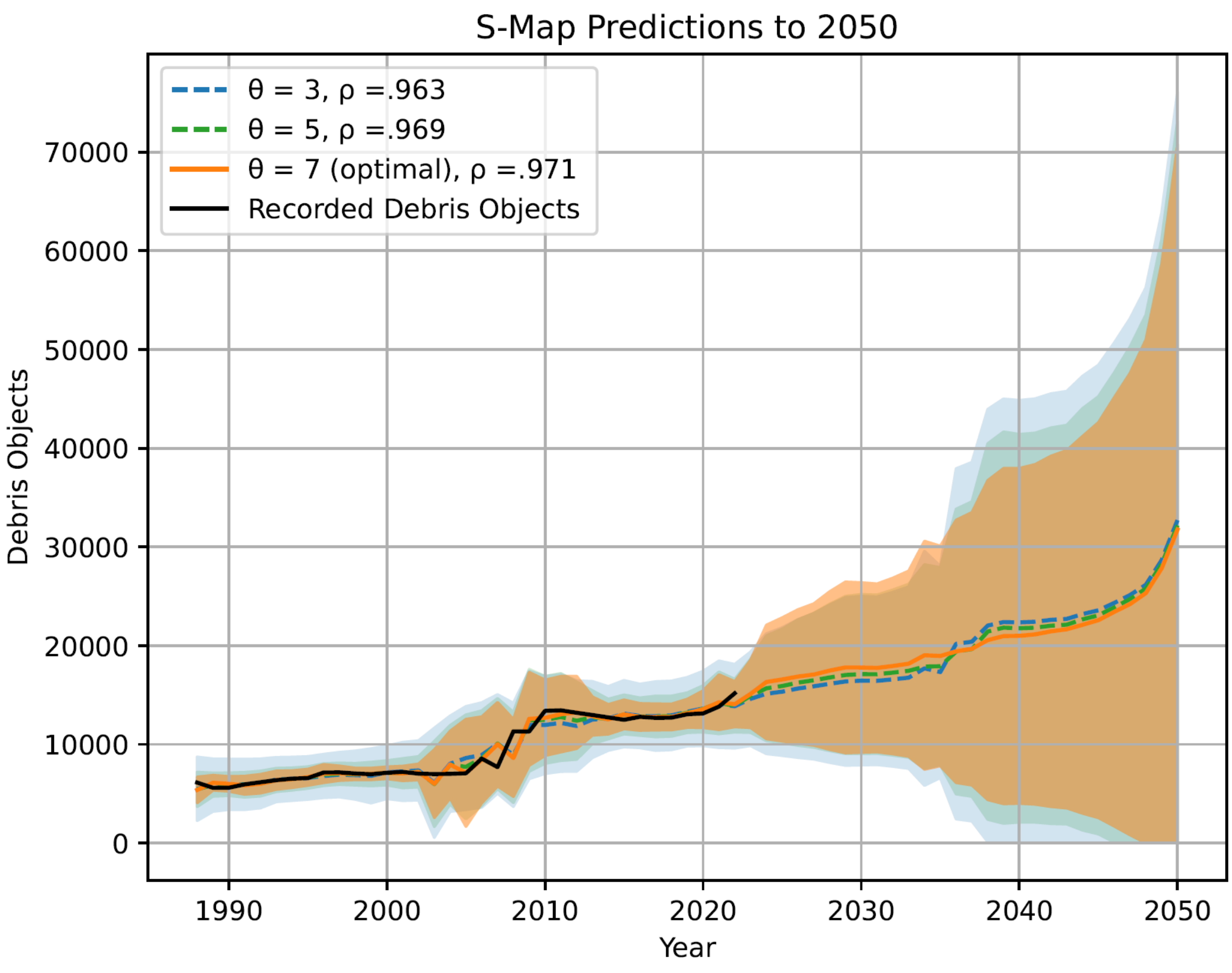

**Fig. 4 S-Mapping forecast out to 2050, with multiple different θ.**

This data-driven method of prediction has different results than parametric predictions commonly found in the literature. Figure 7 features the result of a 25,000-run Monte Carlo simulation using the DAMAGE debris evolutionary model [45]. Shaded area around the mean represents the results of all 25,000 Monte Carlo runs. S-Map results with 95% confidence band are featured for comparison. Parameters for this simulation include 90% compliance with a 25-year post-mission disposal rule, assumptions regarding launch traffic, no collision avoidance maneuvers, and no explosions. For reference, we also included an S-Map prediction based on a 90% compliance rate. This is calculated by using the data for the yearly number of objects launched into space, assuming that each object has a 10-year operational lifespan and a 25-year post-mission disposal period. This 90% compliance rate does not significantly change the results, differing from the primary prediction by less than 0.1%.  Our primary S-Mapping result predicts 13,854 more debris objects than the mean of the Monte Carlo simulation, which is reflective of more dynamics contributing to the debris system's evolvement. Particularly, the inclusion of explosions is a significant contributor.

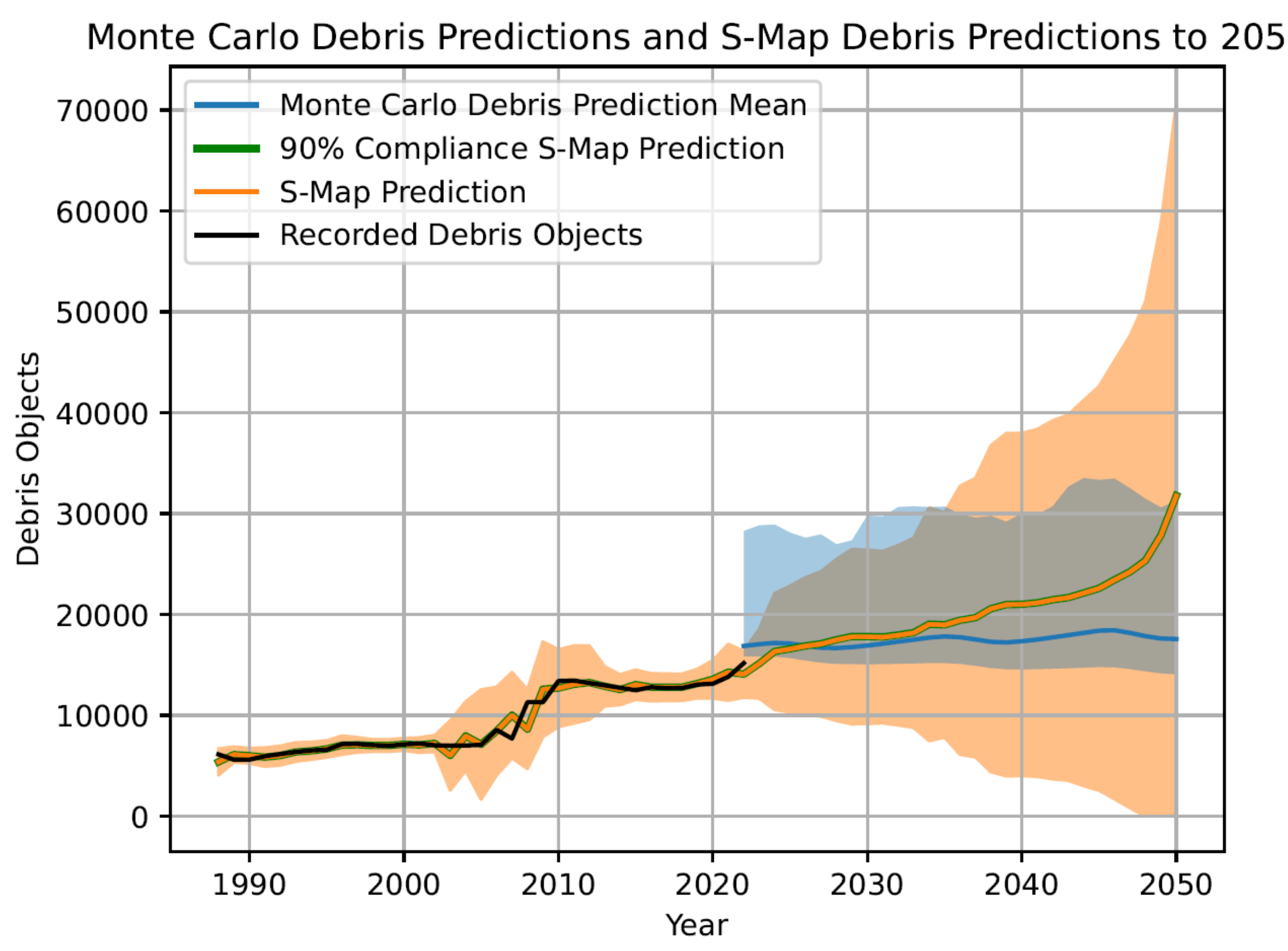

**Fig. 5 Monte Carlo simulations of the evolution of objects > 10 cm in LEO based on the DAMAGE model.**

With the optimal localization parameter θ, S-Mapping can also be performed with different combinations of variables. The S-map prediction for univariate, 2-input multivariate, and full 3-input multivariate configurations can be seen in Fig. 8. Each prediction features a 95% confidence band based on the variance at each prediction point. Univariate considers debris in LEO data (X), 2-input includes combinations of debris in LEO (X) and objects launched into space (Y) or debris in LEO (X) and total bodies in LEO (Z) data, and the full multivariate considers data from debris in LEO (X), objects launched (Y), and total bodies in LEO (Z).

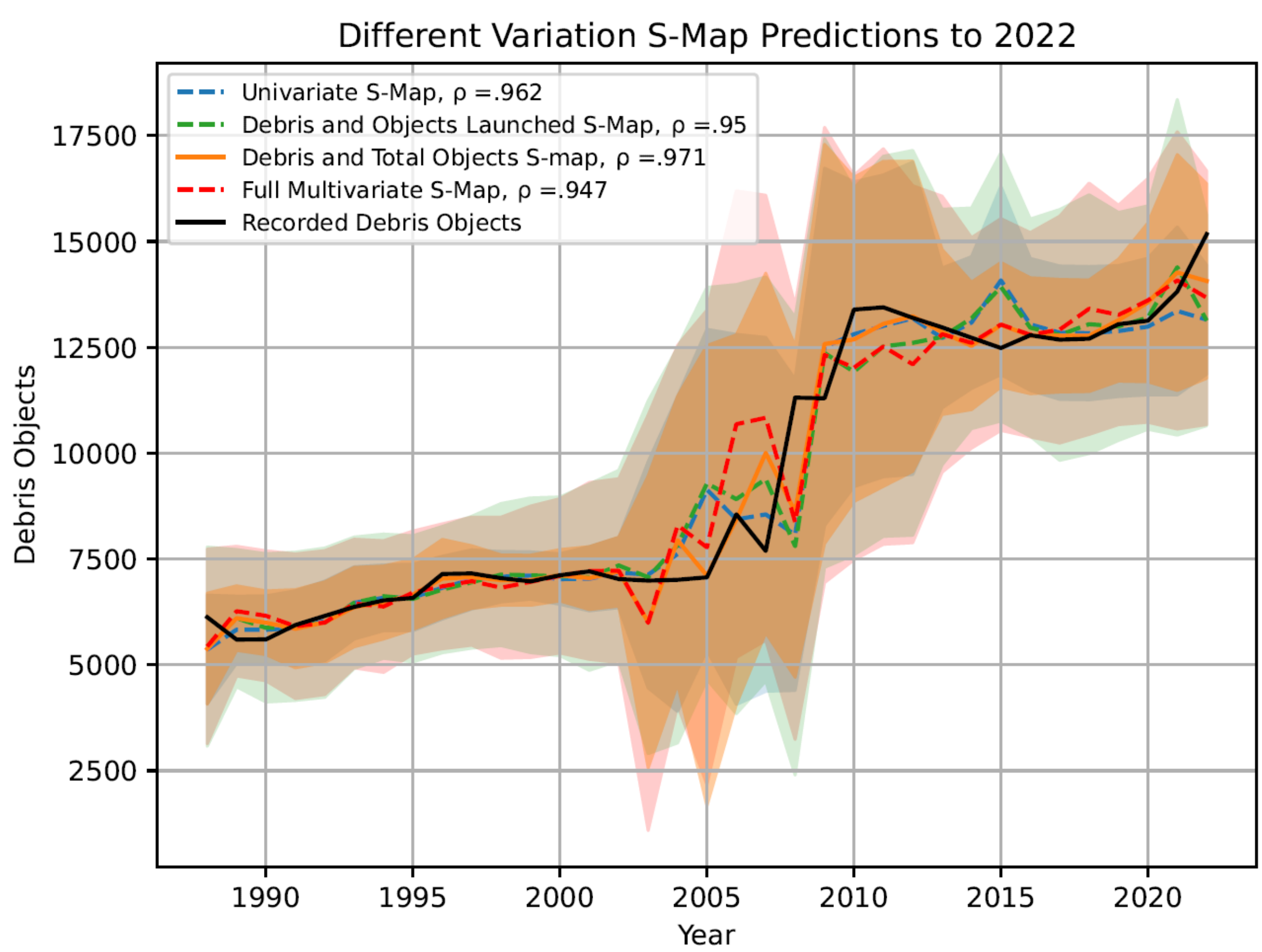

**Fig. 8 S-Mapping Predictions using univariate, 2-input multivariate, and 3-input full multivariate.**

These results indicate that the highest prediction skill is available by considering the X and Z variables, which corresponds with the results from our CCM analysis. Using combinations of variables which have weaker causal relationships diminishes the prediction skill, as seen in S-Map predictions using X and Y or all three variables. These results also follow the behavior that the prediction with the highest prediction skill experiences the least uncertainty.

### E. Policy Simulation

Regulatory bodies are beginning to implement policies to mitigate the creation of orbital debris. In 2022 the United States' Federal Communications Commission (FCC) updated their post-mission disposal (PMD) time policy for non-geostationary satellites and space stations to be from 25 years to 5 years [46]. The goal of this policy is to mitigate debris generation by reducing the amount of post-mission spacecraft that could be involved in debris generating events. These kind of policy decisions are currently supported by a limited amount of quantitative data on how they will actually impact the orbital debris environment.

EDM can be used to provide a specific quantitative basis on how a policy change will impact the rest of the orbital debris system. We consider three simulation scenarios in this research, all of which are considered to begin being effective in the year 2000. The first is where we assume that a certain number of debris is prevented by limiting post-mission lifetime. By utilizing the annual data of objects launched into space up to 2022, we identify exactly how many launched objects should be de-orbited each year based on a given PMD policy. We assume each object has an operational lifetime of 10 years, after which the PMD policy comes into effect. The 15-year PMD policy is effective up to 2047, the 10-year PMD policy is effective up to 2042, the 5-year PMD policy is effective up to 2037, and the 0-year PMD policy is effective up to 2032. S-map was used to predict the number of debris objects and the number of total bodies in LEO up to 2050. Then, for the time range of each PMD policy, the number of debris objects and total bodies in LEO is subtracted by the amount of objects which should be deorbited. Finally, S-map is used again to predict the remaining years between the PMD policy range and 2050. The results of this simulation allow us to have a relative scale for how the orbital debris environment might change based on what the post-mission disposal lifetime is.

The second simulation scenario considers alternative policies for mitigating orbital debris. We consider a policy that would ultimately reduce the number of objects launched each year by 20%, 15%, 10%, or 5% from their recorded historical number from 2000 to 2022. S-map is then used to predict the altered trajectory from 2023 to 2050. The third simulation considers a series of Active Debris Remediation (ADR) missions that results in the removal of 100, 300, 1000, or 3000 random pieces of debris every year from 2000 to 2022. S-map is again used to predict the altered trajectory from 2023 to 2050. The lower bound scenario of 100 was selected based on NASA's Cost and Benefit Analysis of active debris remediation [11], while the following scenarios were selected to simulate more intense remediation activity. Table 1 contains all assumptions used for policy simulations. Select simulations can be found in Fig. 9, while the full tabulated results of each simulation can be found in Table 2. The select simulations in Fig. 9 are the least and most intensive mitigation measures for each policy simulation. Each prediction features a 95% confidence band based on the variance at each prediction point. The last graph in Fig. 9 illustrates the percentage difference in debris population between the baseline S-map prediction from Fig. 6 and our simulated results. In the case of the Objects Launched Policy simulation, a separate baseline was established since doing analysis with all three variables instead of just X and Z results in a completely different system evolvement. The separate baseline is represented as the 0% Objects Launched mitigation policy.

**Table 1 Simulation Assumptions**

| Mitigation Policy | Current | Assumption |
|---|---|---|
| Post-mission disposal (PMD) policy | 25 Year | 15 Year, 10 Year, 5 Year, 0 Year |
| Active Debris Removal per year | 0 | 100, 300, 1000, 3000 |
| Reduce objects launched | 0% | 10%, 20%, 30%, 40% |

**Table 2 Mitigation Policy Effectiveness**

| Mitigation Policy | S-Map Debris Prediction to 2050 | Margin of Error | Percent of Debris Mitigated in 2050 |
|---|---|---|---|
| 25 Year PMD (Current) | 31,758 | 39,030 | 0% |
| 15 Year PMD | 22,262 | 4,903 | 29.90% |
| 10 Year PMD | 15,514 | 7,793 | 51.14% |
| 5 Year PMD | 18,018 | 3,172 | 43.26% |
| 0 Year PMD | 5,138 | 20,901 | 83.82% |
| 0% Less Objects Launched | 62,156 | 100,194 | 0% |
| 10% Less Objects Launched | 59,139 | 94,286 | 4.85% |
| 20% Less Objects Launched | 56,101 | 88,339 | 9.74% |
| 30% Less Objects Launched | 53,056 | 82,379 | 14.64% |
| 40% Less Objects Launched | 50,004 | 76,409 | 19.55% |
| 100 ADR Policy | 31,534 | 38,775 | 0.67% |
| 300 ADR Policy | 31,126 | 38,283 | 1.99% |
| 1000 ADR Policy | 29,755 | 36,743 | 6.31% |
| 3000 ADR Policy | 26,489 | 33,684 | 16.59% |

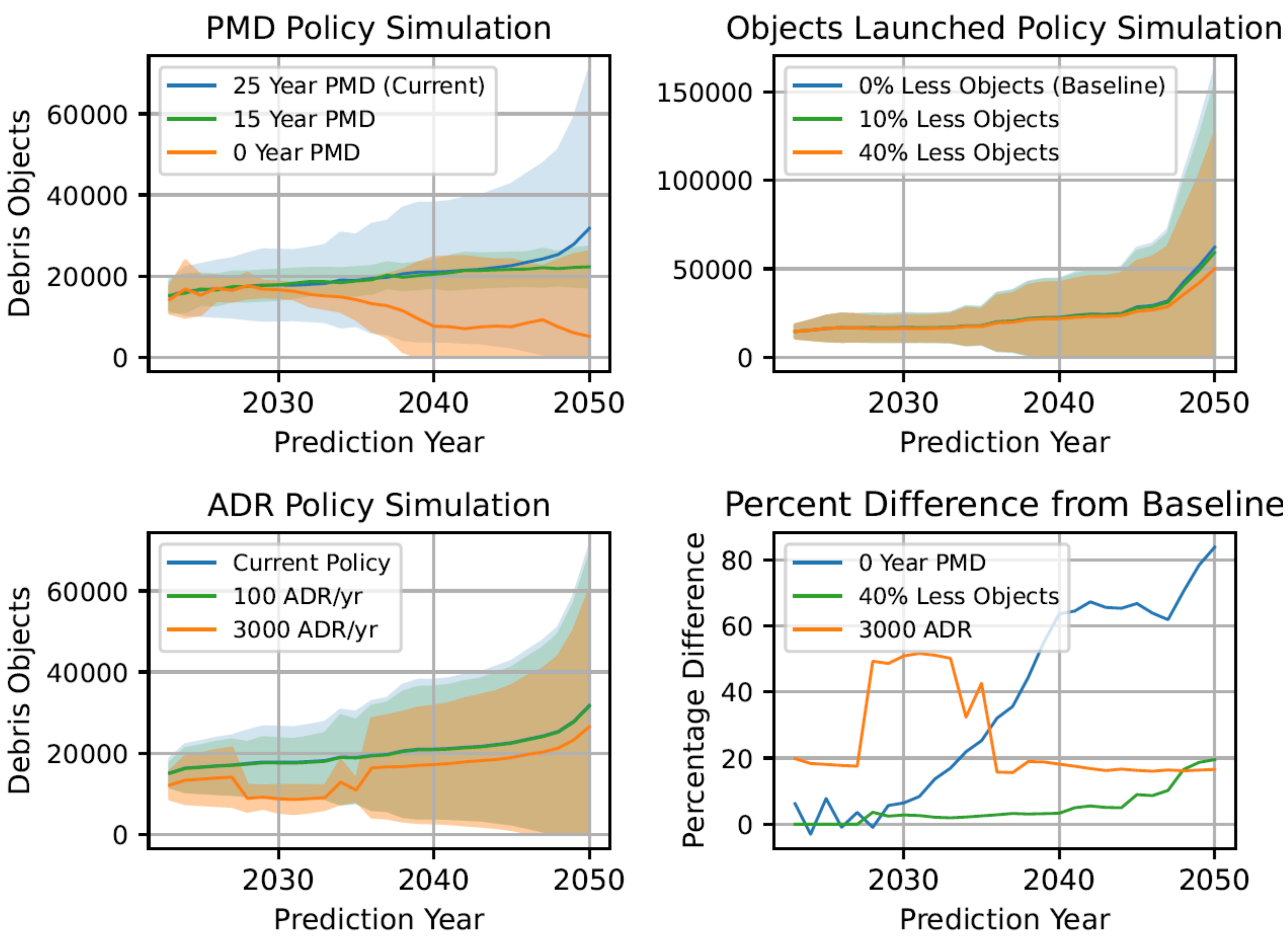

**Fig. 9 Simulations of Orbital Debris based on policy changes effective in 2000.**

The PMD policy simulations in Fig. 9 feature confidence bands that vary greatly, and this is a result of the separate prediction horizons for the final S-Map in each prediction scenario. Since the 15-year PMD scenario only has a final prediction horizon of 3 years, and the 0-year PMD scenario only has a final prediction horizon of 18 years, there is less time for the uncertainty to expand when compared to the 28-year baseline prediction horizon. The policy simulations for reducing the annual number of objects launched exhibit a higher predicted debris population than the baseline from S-Mapping with X and Z variables. This is because the objects launched policy simulation also includes the Y variable as a system input. This results in not only lowered prediction accuracy, as discussed in the S-Map section, but also drastically changed prediction trajectory. The objects launched policy simulation provides a significant result of nearly 20% debris objects mitigated. While this is substantial, the growth of the space industry, especially regarding the rise of mega-constellations, presents conflicts to practically utilizing this mitigation method. The ADR policy simulation for 3000 debris objects removed per year features an interesting trend, where the population drops significantly in 2027, but returns to a trajectory similar to the baseline by 2035. This behavior is consistent with the prior implementation of the ADR into our simulated historical data. In this scenario, the S-Map prediction has calculated that intense ADR would happen again from 2027 to 2035, which is the result of our simulation introducing a clear dynamic into the system's evolvement.

The metric of success for these mitigation policies is the percentage difference in predicted debris population in 2050, as seen in Table 2, and the bottom right plot of Fig. 9. One notable result is that 10-year PMD policy is more effective than the 5-year PMD policy. It is unintuitive that the less intense mitigation policy would have a greater impact on reducing the debris population. This result is caused by the cadence of deorbiting for the 10-year PMD policy being weighted more heavily in the S-Mapping, since the timing of a change to empirical data can be more important than the quantitative magnitude of that change. The higher margin of error for the 10-year PMD scenario and 0-year PMD scenario indicate that the system dynamics in those policy simulations are more difficult to capture. Another interesting trend from Table 2 is that for the non-PMD simulations, which share the same prediction horizon, the margin of error decreases as the mitigation policy becomes more intense. This indicates that the artificial system dynamics become easier to capture as they become more intense, since their impact on the system becomes more obvious.

Results indicate that the most effective simulated policy was reducing post-mission lifetime, with reducing objects launched and active debris remediation methods falling significantly behind. Given the high demand for increased space operations, a trend that shows no sign of decreasing, reducing the number of object launches would be the least realistic mitigation strategy. Active debris remediation is a method that would improve with more servicers to aid the debris remediation missions, so it stands to become more viable as those capabilities are developed further into the future. The post-mission disposal lifetime policy is most viable for the near and far future, since the launch of satellite mega-constellations in recent years has drastically increased the number of total bodies in LEO.

The results of these simulations reflect the differences between our policy simulation methodology and other policy simulation methods in the literature. PMD policy simulation typically considers a 90% compliance rate for PMD success [2], whereas we consider a 100% success rate. While this allows us to display the maximum possible effectiveness of PMD policy changes, in reality the effectiveness would fall short of these estimations. Additionally, ADR typically considers specific debris objects to be remediated, not random objects as we have selected. These debris objects are typically identified based on their likelihood of increasing the debris population based on size, probability of collision, and probability of explosion [2,3,11]. A limitation of our data-driven method is that we cannot account for weighing debris objects differently. This significantly changes the ultimate effectiveness of ADR implementation, as removing the most dangerous objects would have a greater impact on reducing the orbital debris population. Another limitation of our methodology is that it can only capture system dynamics that have been exhibited in the past of time series data. This means that any new emergent dynamics, such as the rapid increase in objects launched into space via mega constellations, are not taken into account until they have time to reflect in the empirical data. Finally, our simulations do not consider composite methods of mitigation, and thus does not explore potential breakpoints in combining ADR, shorter PMD policy, and launching less objects per year.

## IV. Conclusion

Orbital debris poses a significant problem to the future of space operations. Any activity regarding orbital debris is inherently connected with a swathe of other space systems, and there are many distinct actors operating in space who pursue a mix of independent and collaborative interests. The data-driven method proposed in this paper seeks to add another tool to address orbital debris as a system thinking problem. This paper has shown that this framework can analyze and quantify relationships between space system variables, as well as predict future values of any particular system variable, with only limited time series data available. This application will be extremely valuable in testing possible solutions to orbital debris, both in active debris remediation and preemptive policy changes, all without the use of equation-based modeling.

The primary constraint for this research was the time series data available. As EDM only utilizes time series data, selection of data is paramount to achieving significant results. Having more sets of time series data that relate to the orbital debris problem would be helpful for developing our ability to understand the impact and nature of the orbital debris system. For instance, time series data on the number of debris generating events or the annual mass of objects launched into space would be useful. Space operators in the public and private sectors, who have access to more detailed or proprietary information about space activity, would be well served by utilizing EDM. Accessing a wider breadth of time series system variable data improves the accuracy with which we can reconstruct the larger system.

The future of work with this method may focus on analyzing a larger array of space systems. This paper focused on the direct future of the orbital debris environment, but there are other space systems which also bear analysis. The economic, environmental, and policy impacts of orbital debris solutions can also be analyzed and simulated. The limitation in analysis via time series data is that one must ensure that there are enough data points which contain the characteristic behaviors of a system. The simulation utilized in this paper was simplified due to the scope of available data. A more sophisticated method of relating policy change to time series data is worth investigating.

## V. Acknowledgments

This material is based upon work supported by the National Science Foundation under Grant No. 2301627. Any opinions, findings, and conclusions or recommendations expressed in this material are those of the authors and do not necessarily reflect the views of the National Science Foundation.